\documentclass[doublecol]{epl2}

\usepackage{graphicx, color}
\usepackage{amsmath,amssymb}
\usepackage[english]{babel}
\usepackage[utf8]{inputenc}
\usepackage{cite}

\title{Geometric signature of complex synchronisation scenarios}

\author{J.H. Feldhoff\inst{1,2}\thanks{E-mail: \email{feldhoff@pik-potsdam.de}}, R.V. Donner\inst{1}, J.F. Donges\inst{1,2}, N. Marwan\inst{1} \and J. Kurths\inst{1,2,3}}
\shortauthor{J.H. Feldhoff \etal}

\institute{
  \inst{1} Potsdam Institute for Climate Impact Research - P.O. Box 60 12 03, 14412 Potsdam, Germany\\
  \inst{2} Department of Physics, Humboldt University - Newtonstr. 15, 12489 Berlin, Germany\\
  \inst{3} Institute for Complex Systems and Mathematical Biology, University of Aberdeen, Aberdeen AB24 3FX, United Kingdom
}

\abstract{
Synchronisation between coupled oscillatory systems is a common phenomenon in many natural as well as technical systems. Varying the strength of coupling often leads to qualitative changes in the complex dynamics of the mutually coupled systems including different types of synchronisation such as phase, lag, generalised, or even complete synchronisation. Here, we study the geometric signatures of coupling along with the onset of generalised synchronisation between two coupled chaotic oscillators by mapping the systems' individual as well as joint recurrences in phase space to a complex network. For a paradigmatic continuous-time model system, the transitivity properties of the resulting joint recurrence networks display distinct variations associated with changes in the structural similarity between different parts of the considered trajectories. They therefore provide a useful indicator for the emergence of generalised synchronisation. \\
\\
\emph{This paper is decidated to the 25th anniversary of the introduction of recurrence plots by Eckmann et~al.~(EPL, \textbf{4} (1987), 973).}}

\begin{document}

\maketitle

\section{Introduction}

The scientific interest in studying and qualitatively understanding synchronisation between coupled oscillatory systems traces back at least to observations of weakly coupled pendulum clocks made by Christiaan Huygens in the 17th century~\cite{Pikovsky2001,Boccaletti2002}. In the last decades, a detailed quantitative knowledge of synchronisation phenomena has been established for both periodic and chaotic oscillators~\cite{Fujisaka1983,Kuramoto1984,Afraimovich1986,Pecora1990,Rosenblum1996}. 

In the study of synchronisation processes between coupled chaotic oscillators, different phenomena have to be distinguished: First, complete synchronisation (CS)~\cite{Pecora1990} can occur for identical systems with a sufficiently strong coupling. In this case, the trajectories $x(t)$ and $y(t)$ of the two coupled systems become identical, i.e. $y(t)=x(t)$. Second, generalised synchronisation (GS)~\cite{Rulkov1995,Abarbanel1996,Kocarev1996,Boccaletti2000a} refers to a general fixed and deterministic functional relationship between both trajectories, $y(t)=f(x(t))$, where $f$ is a diffeomorphism. This type occurs for non-identical chaotic oscillators at rather high coupling strengths. A closely related form is lag synchronisation (LS)~\cite{Rosenblum1997}, where both coupled systems evolve in an identical way with a fixed mutual time shift as $y(t)=x(t-\tau)$. Finally, phase synchronisation (PS) is characterised by a locking between the phases of two systems, $m\phi_y(t)=n\phi_x(t)$ with $m,n\in\mathbb{N}$, whereas the amplitudes may remain uncorrelated. PS is typical for non-identical chaotic systems at moderate or even weak coupling strengths.

Even though PS is typically considered a weaker form of synchronisation than GS and CS, the sequence of synchronisation phenomena arising for two coupled oscillators with increasing coupling strength depends crucially on the structural differences between both systems as well as the specific coupling configuration. For several examples, it has been demonstrated that as the coupling becomes stronger, the dynamics of both systems becomes successively synchronised on more and more characteristic time-scales~\cite{Hramov2004,Hramov2005}. This observation implies the emergence of PS at relatively low coupling strengths if the characteristic oscillation frequencies of both systems are sufficiently similar, whereas LS, GS, and CS typically occur at higher coupling strengths. However, under certain conditions the observed synchronisation scenario can be much more complex.

A formal description of GS has been given first only for driver-response systems. However, the typical signatures of GS are not unique to systems with unidirectional coupling, but can be observed in a similar way also for symmetrically coupled oscillators~\cite{Zheng2002PRE} (in fact, GS has been first described for two systems with bidirectional coupling~\cite{Afraimovich1986}). 

For driver-response relationships, the driven system undergoes systematic changes in its dynamical evolution as the coupling strength is increased. These changes can be detected by a variety of methods such as the auxiliary system method~\cite{Abarbanel1996}, numerical estimates of the conditional Lyapunov exponent of the response system as a measure for its asymptotic stability~\cite{Kocarev1996}, or by quantifying the predictability of the driven system. The latter class of approaches includes several methods based on neighbourhood relationships in phase space (i.e., relying on geometric information), with the mutual false nearest neighbour method~\cite{Rulkov1995} and the synchronisation likelihood~\cite{Stam2002} as most prominent examples. However, there are specific cases (such as systems exhibiting multistability) where at least some of these methods are not applicable. Even more, they are not always directly transferable to the case of bidirectional coupling configurations. Consequently, recently several new approaches have been proposed for studying GS for both uni- and bidirectional couplings, including methods based on recurrence plots~\cite{Romano2005EPL} or a generalised angle between the trajectories of both systems~\cite{Liu2009EPL}.

Since synchronisation has distinct effects on the spatial organisation of driven systems in phase space, geometric methods are promising and widely applicable candidates for detecting GS from time series data. In this work we propose a new geometric method based on a complex network approach. Specifically, our method is based on a graph-theoretical interpretation of joint recurrence plots describing the simultaneous occurrence of mutually close pairs of state vectors in two or more coupled systems~\cite{Romano2004PLA}. After presenting the details of our approach, the paradigmatic example of two coupled R\"ossler oscillators in different dynamical regimes are discussed in order to demonstrate the performance of the proposed method. The obtained results indicate that the complex network-based approach has great potentials and may be applied to real-world problems in future work.

\section{Methodology}

In the last decades, there has been a rising interest in studying structural properties of networked systems across disciplines~\cite{Albert2002,Newman2003}. For this purpose, a variety of measures have been introduced that quantitatively characterise different aspects of the observed networks' complex connectivity properties~\cite{Boccaletti2006,Costa2007}. Besides investigating real-world systems with a physical network substrate, recently there have been various successful attempts to applying complex network-based approaches also to more general problems of data analysis, such as for infering so-called functional networks from multivariate time series~\cite{Zhou2006,Donges2009b} or characterising structural properties of time series. For the latter purpose, a multiplicity of complementary approaches (such as cycle networks, visibility graphs, and recurrence networks, to mention only a few) have been proposed~\cite{Zhang2006,Xu2008,Lacasa2008,Yang2008,Marwan2009}, which transform certain types of time series to the complex network domain (see \cite{Donner2011IJBC} for a corresponding review). 

Among the currently existing types of time series networks, approaches characterising the mutual proximity of state vectors in phase space have recently become particularly popular, since the resulting networks' structures reflect nontrivial geometric properties of the supposed (low-dimensional) attractor underlying the observed dynamics. A particularly interesting representative of this class are recurrence networks~\cite{Marwan2009,Donner2010NJP,Donner2011IJBC}, which encode the mutual proximity of state vectors in phase space arising from dynamical recurrences in the sense of Poincar\'e. Specifically, defining neighbourhoods of individual states by considering a spatial threshold distance $\varepsilon$ around each observed state vector offers a flexible way for capturing the geometric backbone of the system under study. Formally, the system's finite-time recurrence properties infered from a time series $\{x_i\}_{i=1}^N$ are then represented by the binary \emph{recurrence matrix}
\begin{equation}
R_{ij}(\varepsilon)=\Theta(\varepsilon-\|x_i-x_j\|),
\label{def:epsrec}
\end{equation}
\noindent
where $\|\cdot\|$ is the maximum norm in phase space (however, other norms could be used here as well), and $\Theta(\cdot)$ denotes the Heaviside function. The visualisation of this symmetric matrix is commonly refered to as the \textit{recurrence plot}~\cite{Eckmann1987} and has become an important tool for nonlinear time series analysis in the last decades~\cite{marwan2007}.

\emph{Recurrence network} (RN) analysis reinterprets matrix (\ref{def:epsrec}) as the adjacency matrix of an undirected simple graph~\cite{Marwan2009} by setting
\begin{equation}
A_{ij}(\varepsilon)=R_{ij}(\varepsilon)-\delta_{ij}
\end{equation}
\noindent
(where $\delta_{ij}$ is Kronecker's delta). The resulting networks' properties can be described analytically by interpreting RNs as random spatial graphs with the system's invariant density uniquely determining RN connectivity~\cite{Donner2011EPJB,Donges2012PRE}. This solid theoretical foundation allows using RNs as a widely applicable tool for detecting dynamical changes in non-stationary time series from mathematical models as well as real-world applications~\cite{Donges2011NPG,Donges2011PNAS,Zou2012bChaos}. One basic, yet important characteristic of RNs is the edge density 
\begin{equation}
\rho(\varepsilon)=\frac{1}{N(N-1)}\sum_{i,j} A_{ij}(\varepsilon),
\end{equation}
\noindent
which is a monotonous function of $\varepsilon$~\cite{Donner2010PRE}. In general, a suitable resolution of small-scale network features requires the choice of a low edge density, typically $\rho\lesssim 0.05$~\cite{Donges2012PRE,Donner2010PRE}, which still meets the condition $\rho>\rho(\varepsilon_c)$ where $\varepsilon_c$ is the percolation threshold of the RN~\cite{Donges2012PRE}.

It has been shown that among others, the transitivity properties of RNs provide some particularly useful characteristics measuring the effective local as well as global dimensionality of the underlying dynamical system~\cite{Donner2011EPJB}. Since in a given integer-dimensional manifold, the transitivity properties can be computed analytically~\cite{Dall2002}, generalising the corresponding relationship to non-integer values allows defining the so-called transitivity dimension
\begin{equation}
D_{\mathcal{T}}(\varepsilon)=\frac{\log\mathcal{T}(\varepsilon)}{\log (3/4)}
\label{eq:dtrans}
\end{equation}
\noindent
with the network transitivity~\cite{Newman2003,Boccaletti2006}
\begin{equation}
\mathcal{T}(\varepsilon)=\frac{\sum_{i,j,k} A_{ij}(\varepsilon) A_{jk}(\varepsilon) A_{ki}(\varepsilon)}{\sum_{i,j,k} A_{ij}(\varepsilon) A_{ki}(\varepsilon)}.
\label{eq:trans}
\end{equation}
\noindent
Other global RN characteristics (e.g., average path length $\mathcal{L}$ or global clustering coefficient $\mathcal{C}$) have also proven their capabilities~\cite{Marwan2009,Donner2011IJBC}, but shall not be further discussed here for brevity. Specifically, the dimensionality interpretation of $\mathcal{T}$ will be the foundation of the approach to studying geometric signatures of synchronisation that will be detailed below.

Motivated by the great potentials of RN analysis of individual dynamical systems, a thorough extension to studying the collective dynamics of two or more coupled systems has been recently proposed in terms of \emph{inter-system recurrence networks (IRNs)}~\cite{Feldhoff2012}. IRNs link the RNs of two different systems using the \emph{cross-recurrence matrix}~\cite{Marwan2002PLA}
\begin{equation}
CR_{ij}(\varepsilon)=\Theta(\varepsilon-\|x_i-y_j\|)
\end{equation}
\noindent
between the associated time series $\{x_i\}$ and $\{y_j\}$. Asymmetries in the resulting cross-transitivity measures for interdependent networks~\cite{Donges2011EPJB} have unveiled geometric signatures of unidirectional couplings between structurally similar systems~\cite{Feldhoff2012}. Note that IRNs do not require simultaneous observations of the considered systems, but imply that they share the same phase space -- a condition which is often not met in real-world applications. Moreover, the application of IRNs is restricted to studying weakly unidirectionally coupled systems before the onset of GS~\cite{Feldhoff2012}.

\begin{figure}[thb]
\centering
\includegraphics[width=0.90\columnwidth]{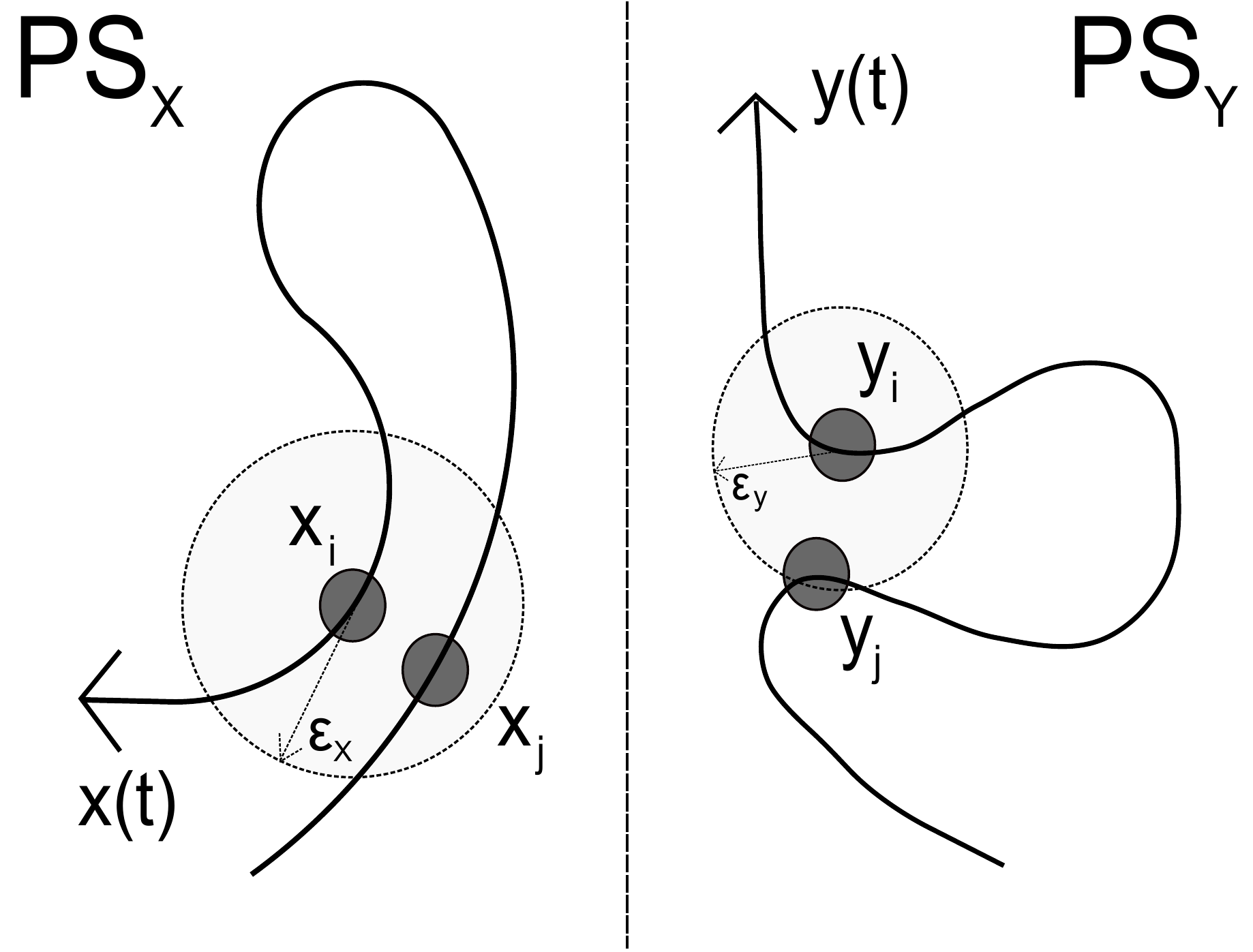}
\caption{Schematic illustration of a joint recurrence of two dynamical systems $X$ and $Y$ in their respective phase spaces $PS_X$ and $PS_Y$.}
\label{fig:joint_rec_scheme}
\end{figure}

In order to study the geometric signatures accompanying the onset of GS in some more detail, we suggest utilising another multivariate generalisation of the recurrence plot concept, \emph{joint recurrence plots}~\cite{Romano2004PLA}, capturing the simultaneous occurrence of recurrences in two dynamical systems (Fig.~\ref{fig:joint_rec_scheme}). The underlying \emph{joint recurrence matrix} 
\begin{equation}
JR_{ij}(\varepsilon_x,\varepsilon_y) = \Theta(\varepsilon_x-\|x_i-x_j\|)  \Theta(\varepsilon_y-\|y_i-y_j\|)
\end{equation}
\noindent
is defined as the point-wise product of the individual systems' recurrence matrices. This basic concept can be directly generalised to the study of $K>2$ coupled systems. 

Next, as for the traditional RN we reinterpret the joint recurrence matrix as the adjacency matrix of a so-called \emph{joint recurrence network} (JRN) by setting
\begin{equation}
A_{ij}(\varepsilon_x,\varepsilon_y)=JR_{ij}(\varepsilon_x,\varepsilon_y)-\delta_{ij}.
\end{equation}
\noindent
Unlike IRNs, JRNs can be constructed for systems with different phase spaces, but explicitly require simultaneous observations of all systems and, thus, time series of the same length. The system-specific thresholds $\varepsilon_k$ should be chosen such that the edge densities $\rho_k(\varepsilon_k)$ of the RNs of the individual systems are fixed at the same value $\rho$. By varying $\rho$, a desired edge density $\rho_J$ can be obtained for the resulting JRN.

The properties of JRNs can be widely interpreted in a similar way as those of a classical RN in the higher-dimensional phase space of the composed system, with the exception that spatial proximity is evaluated separately in the subspaces belonging to the individual systems. Specifically, the transitivity $\mathcal{T}(\varepsilon_x,\varepsilon_y)$ (cf.~Eq.~(\ref{eq:trans})) of the JRN (or, for short, \emph{joint transitivity} $\mathcal{T}_J$) provides a measure for the ``joint dimensionality'' of the composed dynamical system (in a general sense). Due to the larger effective dimensionality of the joint phase space, for two systems $X$ and $Y$ in the absence of synchronisation we expect $\mathcal{T}_X,\mathcal{T}_Y\gg\mathcal{T}_J$ (specifically, $D_{\mathcal{T}_J}=\log\mathcal{T}_J/\log(3/4)\approx D_{\mathcal{T}_X}+D_{\mathcal{T}_Y}$, 
cf.~Eq.~(\ref{eq:dtrans})). In turn, when both systems become synchronised (in the sense of GS), the effective degrees of freedom of both systems become mutually locked, resulting in $\mathcal{T}_J\to \mathcal{T}_X,\mathcal{T}_Y$, i.e., both systems acting collectively as one with the same effective dimensionality as the individual systems. Consequently, studying the ratio between the joint transitivity and some mean value (e.g., arithmetic or geometric mean) of the individual systems' RN transitivities can provide a useful geometric indicator for the emergence of GS. Here, we use the arithmetic mean, since it is never smaller than the geometric one, implying a higher discriminatory power of the \emph{transitivity ratio}
\begin{equation}
Q_{\mathcal{T}}=\frac{\mathcal{T}_J}{(\mathcal{T}_X+\mathcal{T}_Y)/2}
\label{eq:qt}
\end{equation}
\noindent
which we will use as an indicator for GS in the following.

\section{Results}

As a well-studied paradigmatic model exhibiting different types of synchronisation phenomena, we study the performance of JRNs for two non-identical R\"ossler oscillators~\cite{Roessler1976}
\begin{equation}
\begin{split}
 \dot{x_1}&=-(1+\nu)x_2-x_3 \\
 \dot{x_2}&=(1+\nu)x_1+ax_2+\mu_{YX}(y_2-x_2) \\
 \dot{x_3}&=b+x_3(x_1-c) \\
 \dot{y_1}&=-(1-\nu)y_2-y_3 \\
 \dot{y_2}&=(1-\nu)y_1+ay_2+\mu_{XY}(x_2-y_2) \\
 \dot{y_3}&=b + y_3(y_1 - c)
\end{split}
\end{equation} 
\noindent
that are diffusively coupled via their second component as in~\cite{Romano2005EPL}. The detuning parameter $\nu$ accounts for a frequency mismatch of the nonlinear oscillations exhibited by both systems, making them non-identical and, thus, prone to GS. In the following, we will keep it fixed at $\nu=0.02$. In turn, for the characteristic parameters $a$, $b$ and $c$, we will mainly keep the same parameters for both systems throughout all further considerations. Specifically, we study the geometric signatures of the complex synchronisation scenarios arising for two different dynamical regimes with chaotic oscillations: the phase-coherent regime ($a = 0.16, b = 0.1, c = 8.5$) and the non-phase-coherent funnel regime ($a = 0.2925, b = 0.1, c = 8.5$). In all simulations, we integrate the coupled system with step size $h=0.001$ until $T=6,000$, i.e., for 6,000,000 time steps. The first 5,000,000 time steps are discarded to avoid even very long transients, and the remaining data are downsampled by a factor of 200 to obtain RNs and JRNs with $N=5,000$ vertices. For comparative purposes, we also consider the recurrence plot-based indices $CPR$ and $JPR$~\cite{Romano2005EPL}. $CPR$, an indicator for PS, is based on the cross-correlation between the two systems' recurrence-based generalised auto-correlation functions, whereas $JPR$ attempts to identify synchronisation by comparing the systems' RN edge densities $\rho_{X,Y}$ to $\rho_J$, which are conjectured to become very similar in case of GS.

\begin{figure*}[thb]
\centering
\includegraphics[width=0.3\textwidth]{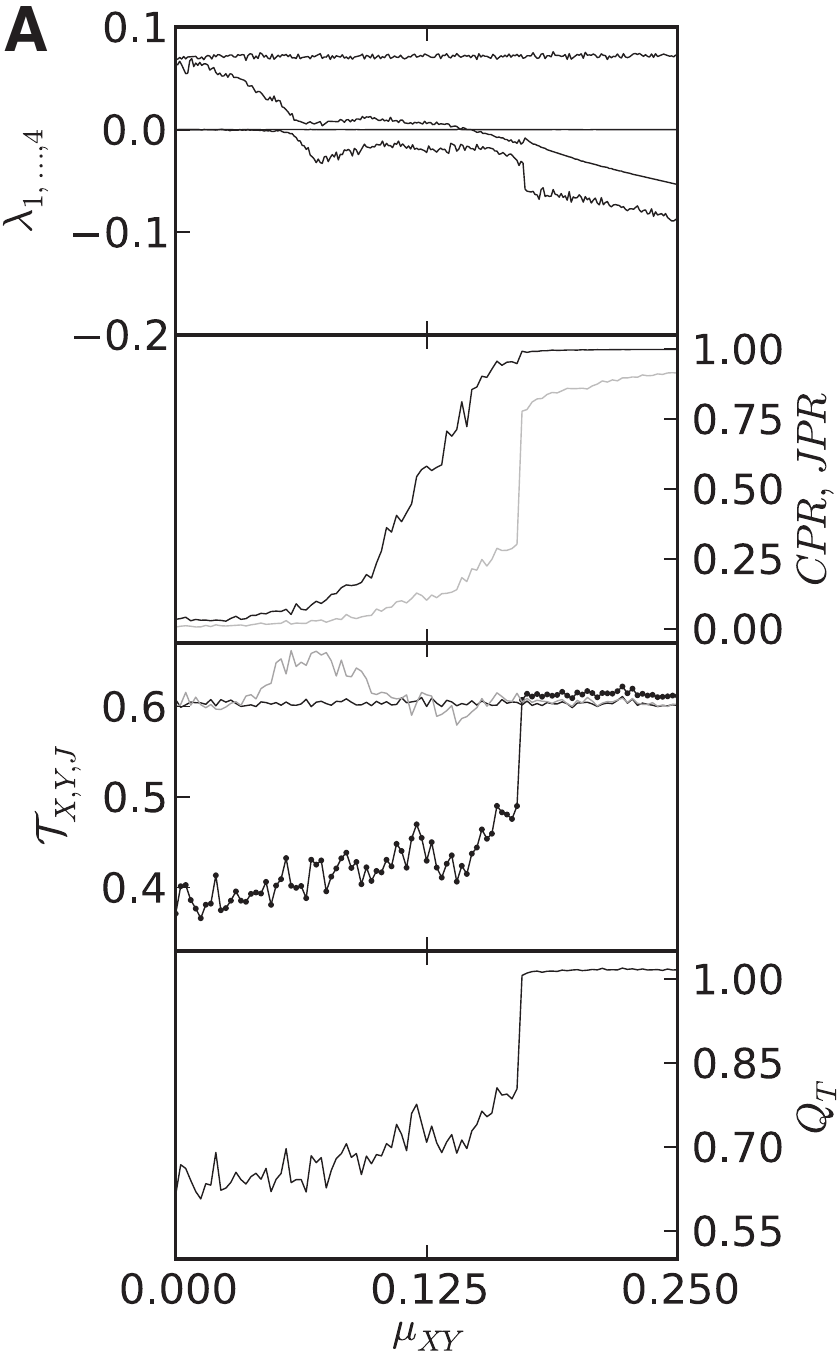}
\includegraphics[width=0.3\textwidth]{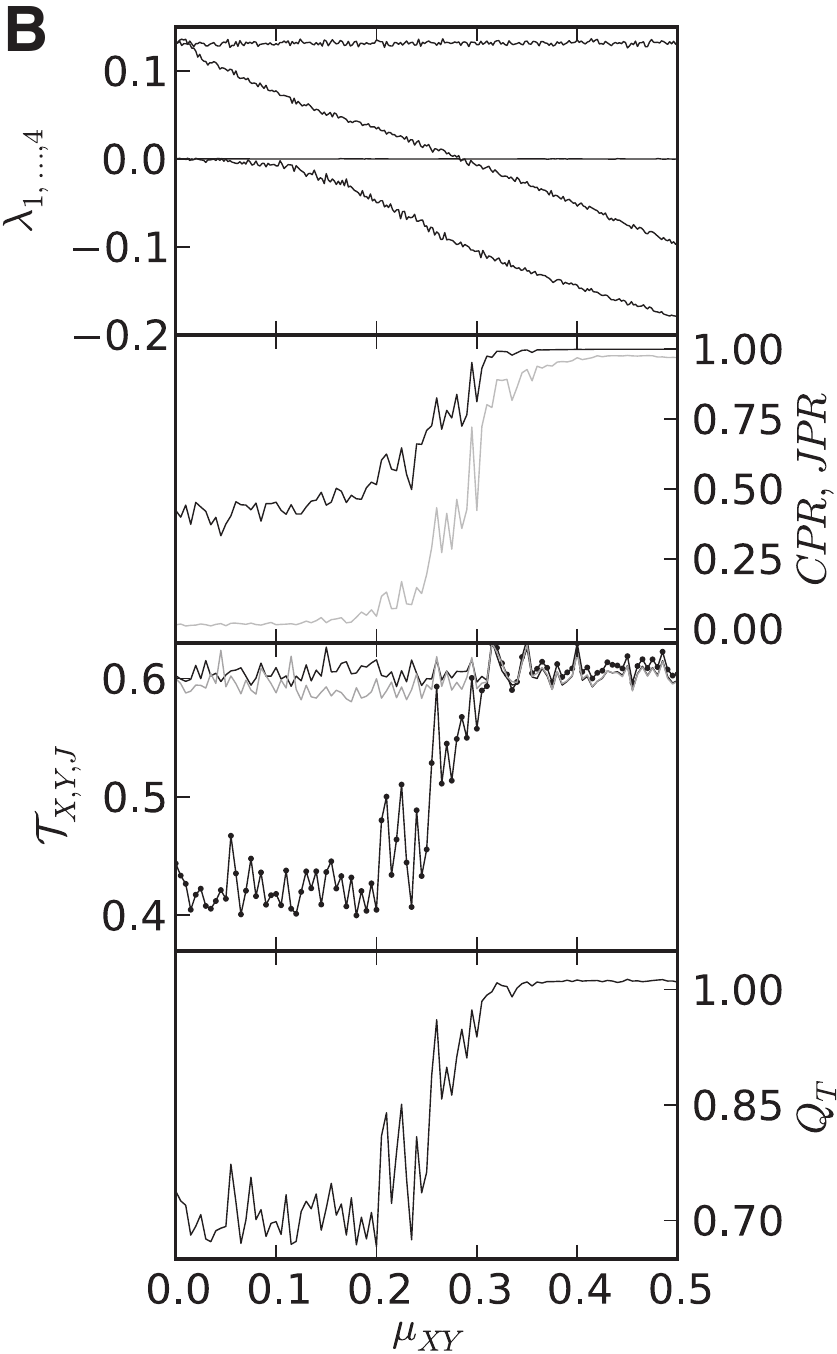}
\includegraphics[width=0.3\textwidth]{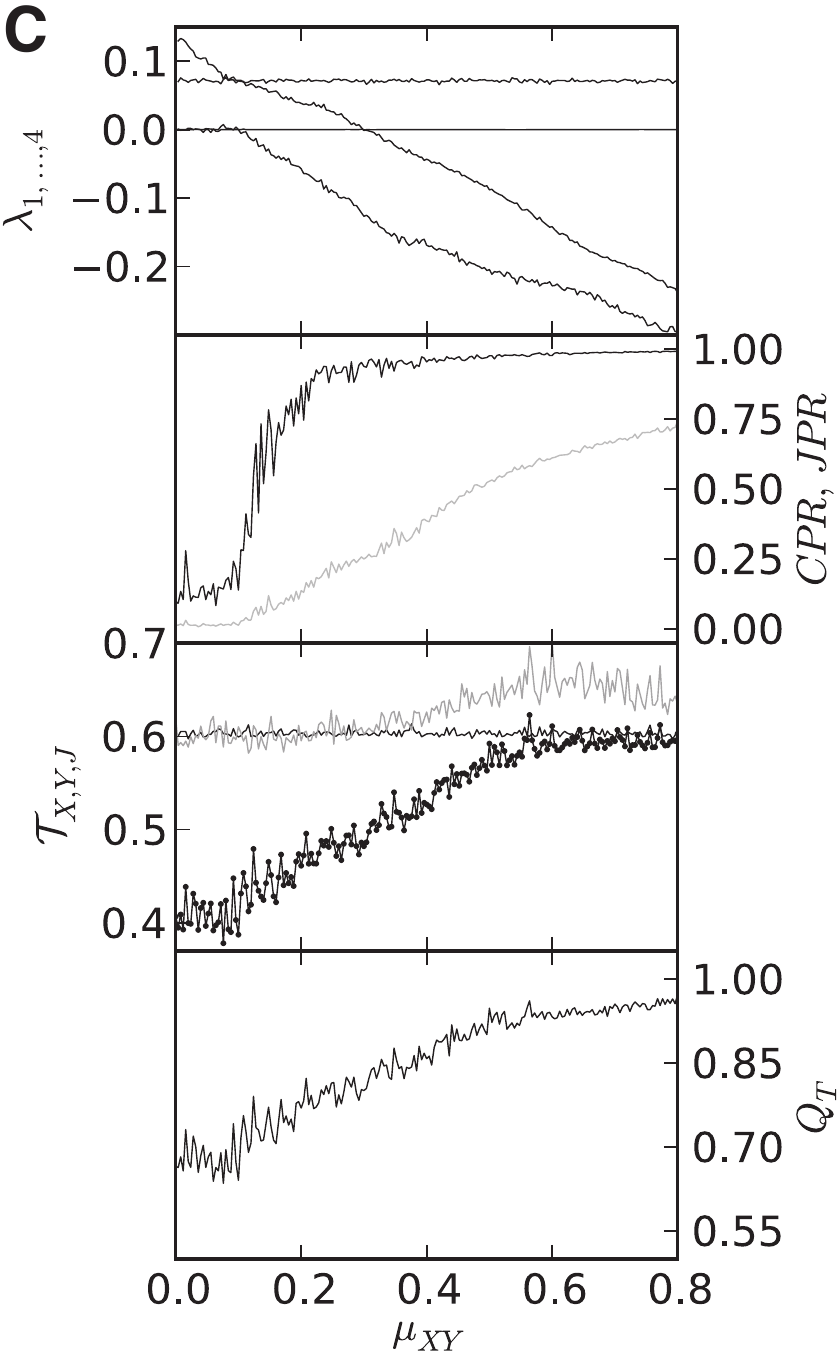}
\caption{Results of synchronisation analysis for two unidirectionally ($X\to Y$) coupled R\"ossler systems being (A) both in the phase-coherent regime, (B) both in the funnel regime, and (C) in phase-coherent ($X$) and funnel regime ($Y$): the four largest Lyapunov exponents $\lambda_1,\dots,\lambda_4$ estimated using the Wolf algorithm~\cite{wolf85} (using $N=9,000,000$ data points from simulations with step size $h=0.001$ starting at $T=1,000$); recurrence-based synchronisation indices $CPR$ (black) and $JPR$ (grey)~\cite{Romano2005EPL}; transitivities of individual and joint recurrence networks $\mathcal{T}_X$ (dark grey), $\mathcal{T}_Y$ (light grey) and $\mathcal{T}_J$ (black); transitivity ratio $Q_{\mathcal{T}}$ (from top to bottom).}
\label{fig:res_unidir}
\end{figure*}

Let us first examine a unidirectional coupling configuration $X\to Y$ ($\mu_{YX}=0$), where the slightly faster system $X$ drives the slower one $Y$. For two phase-coherent R\"ossler systems (Fig.~\ref{fig:res_unidir}A), we find that when increasing the coupling strength $\mu_{XY}$ beyond about 0.06, the fourth-largest Lyapunov exponent $\lambda_4$ gets negative. At about the same value, $CPR$ starts increasing indicating the possible onset of PS. At $\mu_{XY}\approx 0.17$, the third-largest Lyapunov exponent $\lambda_3$ becomes significantly negative, which is commonly interpreted as an indicator for the onset of GS. At the same time, we observe a sharp rise of $JPR$ towards values of about 0.8, whereas $Q_\mathcal{T}$ immediately rises to values close to 1 (in fact, there is some weak overshooting because the edge densities of RNs and JRN differ necessarily, resulting in slightly different estimates of the corresponding network transitivities). This finding suggests that $Q_\mathcal{T}$ has better capabilities for detecting GS than $JPR$.

\begin{figure*}[thb]
\centering
\includegraphics[width=0.3\textwidth]{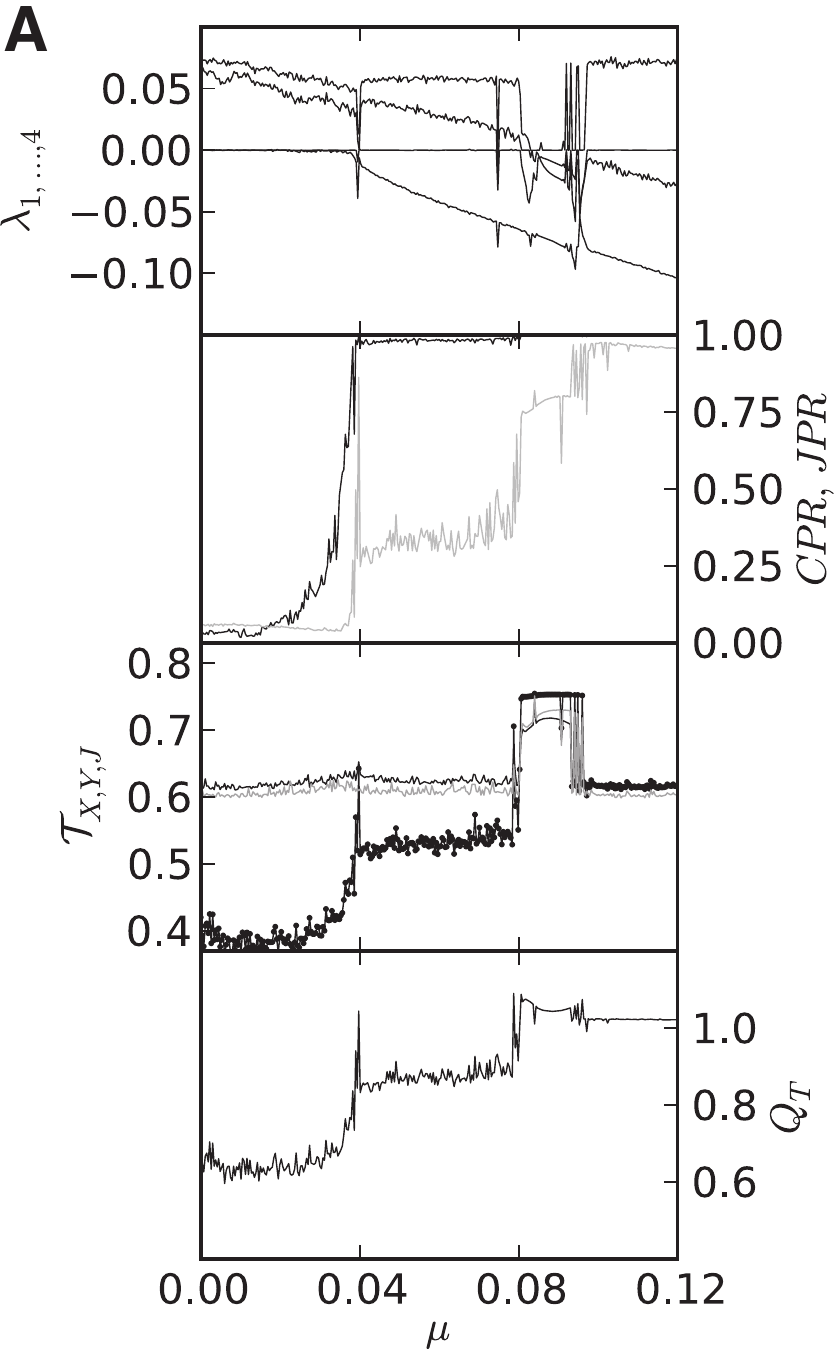}
\includegraphics[width=0.3\textwidth]{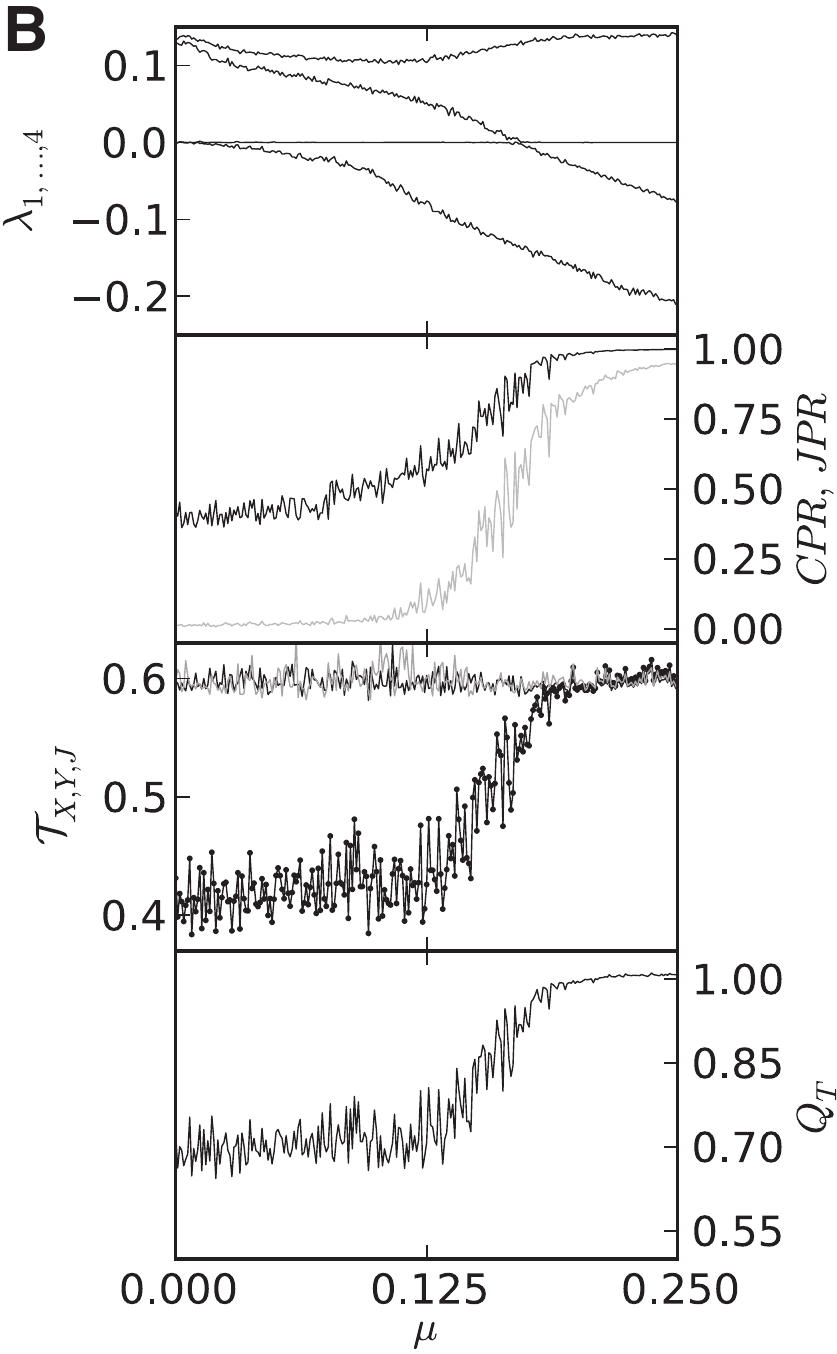}
\includegraphics[width=0.3\textwidth]{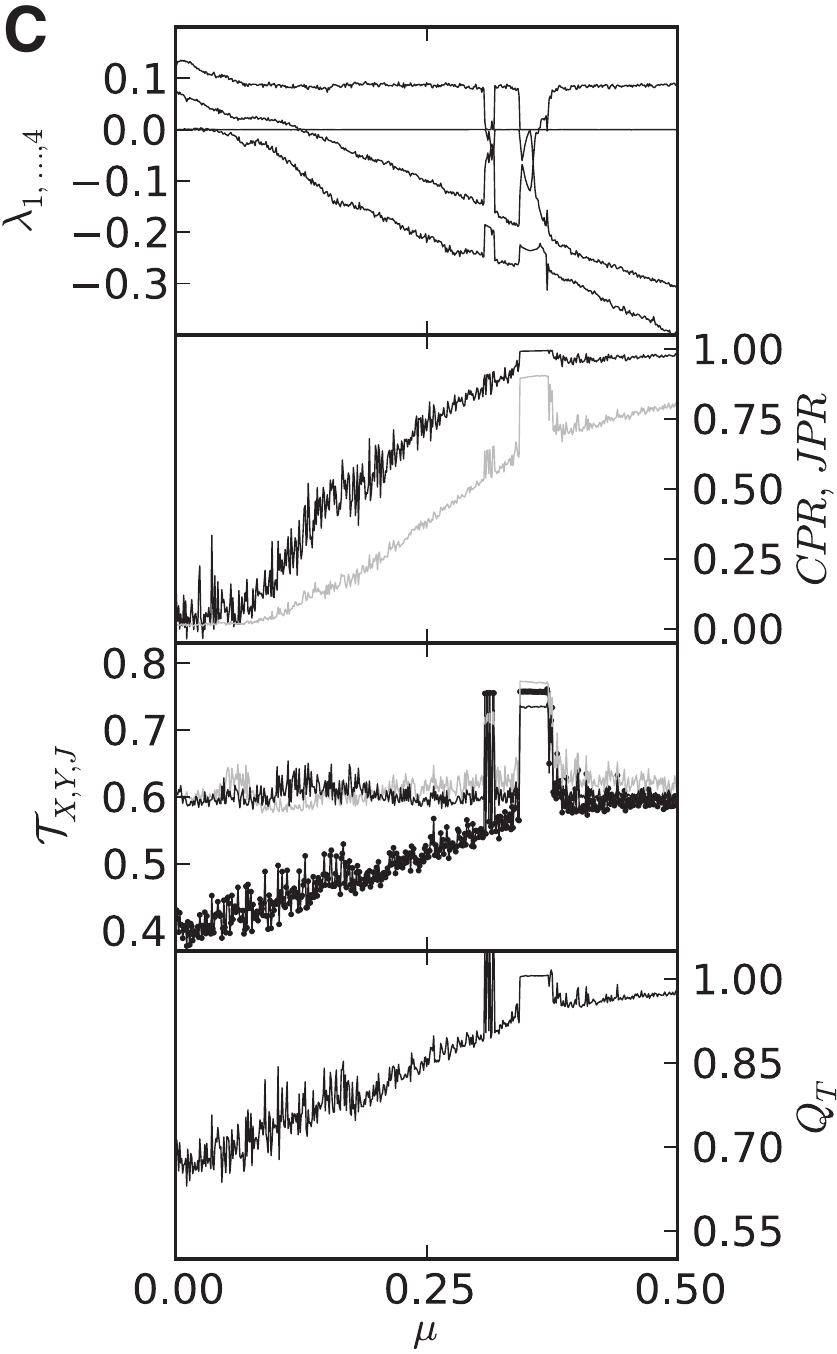}
\caption{As in Fig.~\ref{fig:res_unidir} for bidirectional coupling.}
\label{fig:res_bidir}
\end{figure*}

The results are qualitatively similar when studying the same coupling configuration with both systems being in the non-coherent funnel regime (Fig.~\ref{fig:res_unidir}B). Here, $\lambda_4$ becomes negative already at rather low coupling strengths, whereas $\lambda_3<0$ for $\mu_{XY}\approx 0.3$. The corresponding emergence of GS is accompanied by a continuous transition of $JPR$ and $Q_\mathcal{T}$ to values of about 1, where the JRN-based index again shows better convergence. Moreover, for symmetric bidirectionally coupled funnel systems ($\mu_{XY}=\mu_{YX}=\mu$, Fig.~\ref{fig:res_bidir}B), we observe qualitatively the same behaviour as in the unidirectional case, with the corresponding transitions taking place at smaller values of the coupling strength (i.e., $\lambda_4<0$ for $\mu\gtrsim 0.02$ and $\lambda_3<0$ for $\mu\gtrsim 0.17$~\cite{Romano2005EPL}). Notably, the transition to GS as unveiled by both $JPR$ and $Q_\mathcal{T}$ is even more continuous than for unidirectional coupling, and the residual difference of $JRN$ to the ideal value of 1 close to the transition point is even larger, whereas $Q_\mathcal{T}$ is at the same time already very close to 1.

Studying two symmetrically coupled phase-coherent R\"ossler systems (Fig.~\ref{fig:res_bidir}A), the synchronisation scenario gets more complex due to the emergence of periodic windows as $\mu$ is increased. Specifically, we find a first very small periodic window ($\lambda_1=0, \lambda_2<0$) close to the onset of $PS$ at about $\mu=0.039$, where $CPR$ shows a sharp rise to values close to 1~\cite{Romano2005EPL}. Note that in this as well as other supposedly periodic windows, we observe $\mathcal{T}_X,\mathcal{T}_Y,\mathcal{T}_J\approx 0.75$ consistent with the expected attractor dimension of 1. At $\mu\gtrsim 0.04$, also $JPR$ and $Q_\mathcal{T}$ increase considerably, reaching plateau values in the phase-synchronised chaotic regime~\cite{Romano2005EPL}. For $\mu\in[0.08,0.092]$, we find another more extended window of apparently periodic dynamics $\lambda_2<0$ that probably corresponds to a regime of intermittent lag synchronisation~\cite{Romano2005EPL}, whereas for even larger coupling strength, we infer the presence of GS since $\lambda_3<0$, $JPR\approx 1$ and $Q_\mathcal{T}\approx 1$.

In order to illustrate the limitations of the proposed approach as well as the established $JPR$ index, we finally study the signatures of synchronisation for the mixed case of a phase-coherent R\"ossler system $X$ coupled to a funnel oscillator $Y$, which can be taken as an example for synchronisation between two strongly non-identical systems. In the case of unidirectional coupling (Fig.~\ref{fig:res_unidir}C), we find $\lambda_4<0$ for $\mu_{XY}\gtrsim 0.1$ accompanied by an increase of $CPR$ indicating the presence of PS. At $\mu_{XY}\approx 0.3$, also $\lambda_3$ approaches negative values, whereas both $JPR$ and $Q_\mathcal{T}$ increase gradually, but do not approach values close to 1 even for relatively high coupling strengths. Thus, the three possible GS indicators $\lambda_3$, $JPR$ and $Q_\mathcal{T}$ do not give consistent results in this case, which implies that we cannot unambigously conclude the possible presence of GS here. A similar statement holds for the bidirectional case (Fig.~\ref{fig:res_bidir}C), which displays again a more complicated sequence of transitions.

\section{Conclusions}

25 years after the introduction of recurrence plots by Eckmann \textit{et~al.}~\cite{Eckmann1987}, the development of recurrence-based techniques still continues. With this paper, we would like to honour the seminal work by these authors and show how the paradigms of recurrence plots and complex networks have meanwhile been combined to pave the way for new concepts in complex systems analysis.

We have introduced a new fully geometric approach for detecting GS between two coupled chaotic oscillators based on time series data. Specifically, we have used a complex network interpretation of the recurrence as well as joint recurrence matrices, the transitivity properties of which allow tracing changes in the effective dimensionality of the individual system's attractors as well as the attractor of the combined higher-dimensional system. We have demonstrated that in the presence of GS, both systems effectively behave as one, with the dimensionality of the composed system approaching that of the individual ones.

For the case of structurally similar systems such as slightly detuned chaotic oscillators with otherwise equal characteristic parameters, we have shown that the ratio $Q_\mathcal{T}$ between the JRN transitivity and the arithmetic mean of the individual systems' RNs transitivities quickly approaches a value of 1 as GS takes place. In comparison with the conceptually related (i.e., recurrence plot-based) measure $JPR$, our geometric characteristic $Q_\mathcal{T}$ displays a better convergence towards the value of 1 expected for GS. In turn, for structurally different systems, both indices fail to clearly indicate the transition to GS as unveiled by studying the Lyapunov spectrum.

In general, the proposed method has the advantage of being applicable to relatively short time series of length $N\lesssim\mathcal{O}(10^3)$, whereas the computational requirements for numerically estimating the Lyapunov spectrum are far higher and usually not met in case of real-world applications. In this spirit, we conclude that the proposed method has great potentials for future applications to observational data from different fields of research.

\acknowledgments
This work has been financially supported by the IRTG 1740/ TRP 2011/50151-0 (funded by the DFG/FAPESP), the Leibniz Society (project ECONS), the German National Academic Foundation, and the Potsdam Research Cluster for Georisk Analysis, Environmental Change and Sustainability (PROGRESS, support code 03IS2191B). For calculations of complex network measures, the software package \texttt{pyunicorn} has been used on the IBM iDataPlex Cluster of the Potsdam Institute for Climate Impact Research.

\bibliographystyle{eplbib}
\bibliography{jrn_library}

\end{document}